\documentclass[conference]{IEEEtran}
\IEEEoverridecommandlockouts

\usepackage{cite}
\usepackage{amsmath,amssymb,amsfonts}
\usepackage{algorithmic}
\usepackage{graphicx}
\usepackage{textcomp}
\usepackage{xcolor}
\usepackage{xfrac}
\usepackage{notation}
\usepackage{soul}

\DeclareMathAlphabet{\mathpzc}{OT1}{pzc}{m}{it}

\newcommand{\ist}{\hspace*{.3mm}}
\newcommand{\rmv}{\hspace*{-.3mm}}
\newcommand{\etal}{\textit{et al. }}
\newcommand{\symB}{\text{b}}
\newcommand{\varB}{\sigma^{\symB}}
\newcommand{\varU}{\sigma^{\text{u}}}
\newcommand{\GScalar}{\gamma}
\newcommand{\FB}{\Set{F}^{\symB}}
\newcommand{\BScalar}{\eta}
\newcommand{\PScalar}{\alpha}
\newcommand{\PScalarB}{\PScalar^{\symB}}
\newcommand{\gTransformed}{\underline{g}}
\newcommand{\spacel}{\ist l}
\newcommand{\SetStriation}{\Set{L}}

\newcommand{\PScalarBScaHat}{\hat{\PScalar}^{\symB}}

\newcommand{\rf}{r\text{-}f}
\newcommand{\rr}{\dot{\bm{r}}}

\usepackage{bm}
\def\BibTeX{{\rm B\kern-.05em{\sc i\kern-.025em b}\kern-.08em
    T\kern-.1667em\lower.7ex\hbox{E}\kern-.125emX}}

\DeclareMathOperator*{\argmax}{arg\,max}

\begin{document}

\title{A New Statistical Model for Waveguide Invariant-Based Range Estimation in Shallow Water \\
}

\author{\IEEEauthorblockN{Junsu Jang}
\IEEEauthorblockA{\textit{Scripps Institution of Oceanography} \\
\textit{University of California, San Diego}\\
La Jolla, CA, USA \\
jujang@ucsd.edu}
\and
\IEEEauthorblockN{Florian Meyer}
\IEEEauthorblockA{\textit{Scripps Institution of Oceanography}\\ \textit{Department of Electrical and Computer Engineering} \\
\textit{University of California, San Diego}\\
La Jolla, CA, USA \\
flmeyer@ucsd.edu}
\vspace*{-7mm}
}

\maketitle

\begin{abstract}
Navigation and source localization in the undersea environment are challenged by the absence of a ubiquitous positioning system. Passive acoustic ranging offers a valuable means of obtaining location information underwater. We present a range estimation method based on waveguide invariant (WI) theory, using ship noise recorded by a hydrophone as an acoustic source. The WI is a scalar parameter that describes the interference patterns in spectrograms caused by the interaction of acoustic wave modes propagating in a waveguide, such as shallow water. WI theory enables ranging using a single receiver without detailed knowledge of the environment. In this paper, underwater acoustic signals radiated by a moving large ship, which include broadband and tonal components, are employed for WI-based ranging in a range-independent shallow water environment. In particular, we develop a likelihood function for WI-based range estimation by introducing a statistical model for high signal-to-noise ratio scenarios. Here, the broadband component is assumed to dominate the background noise. The effectiveness of the proposed range estimation method is demonstrated using real acoustic measurements of a moving container ship recorded during the Seabed Characterization Experiment 2017 (SBCEX17).
\end{abstract}

\begin{IEEEkeywords}
Underwater Acoustics, Ranging, Parameter Estimation
\vspace{-3mm}
\end{IEEEkeywords}

\section{Introduction}

\begin{figure}
    \centering
    \begin{minipage}{\linewidth}
        \centering
        \centerline{\includegraphics[scale=.4]{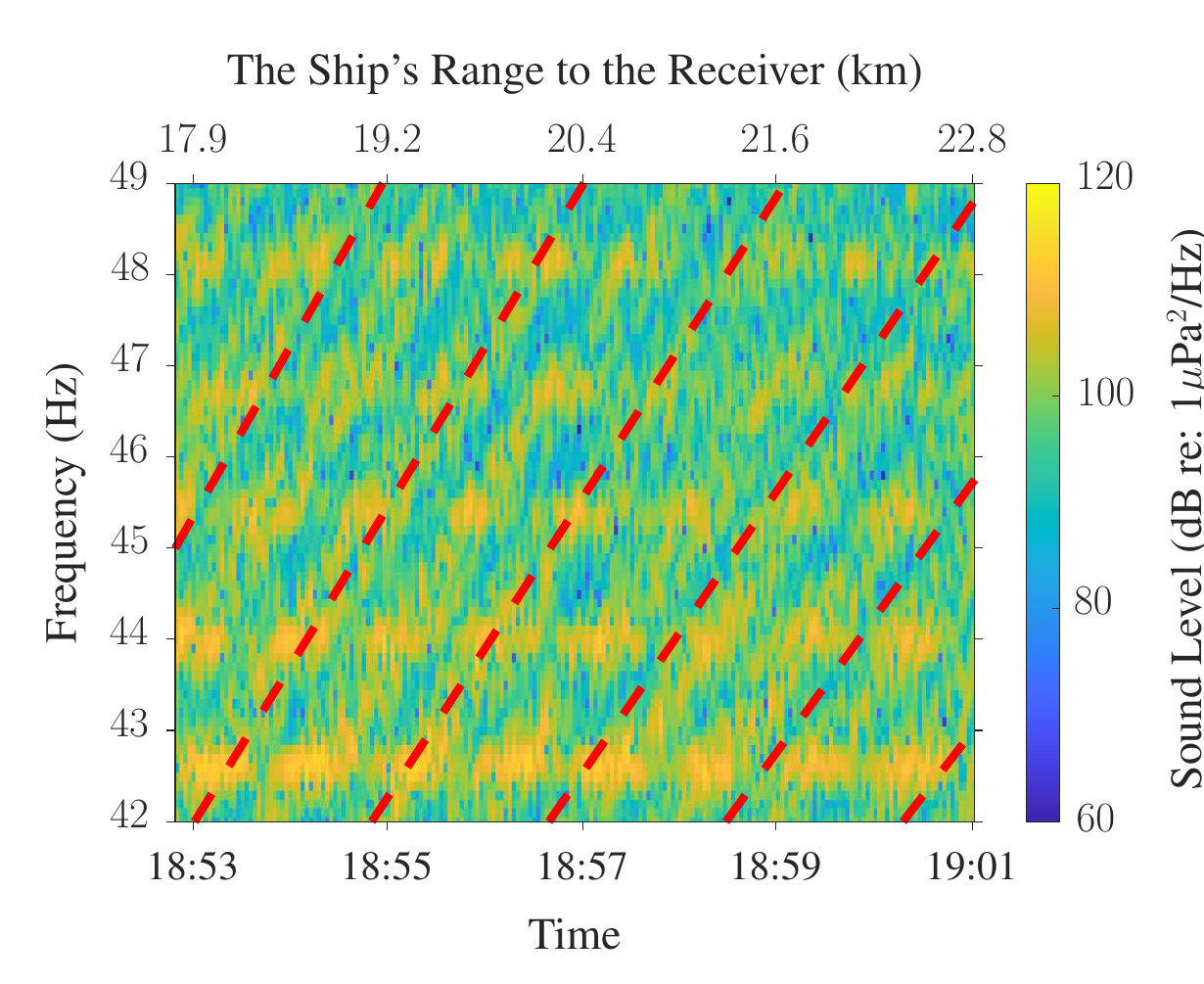}}
    \end{minipage}
    \caption{Spectrogram recorded on March 24, 2017 (UTC) during the Seabed Characterization Experiment 2017 (SBCEX17) using a single element of a vertical line array (VLA). The container ship \textit{Kalamata} passed near the VLA, producing a spatial interference pattern. Five horizontal line spectra (tonals) are clearly distinguishable from the background broadband components emitted by the vessel. The spatial interference is visible across all frequencies, including the broadband frequencies. Red dashed lines represent example striations projected using equation \eqref{eq:wi} and waveguide invariant of $\beta = 1.21$, the established value for this shallow water environment.}
   \label{fig:shipSoundEx}
   \vspace{-4mm}
\end{figure}

The waveguide invariant (WI), denoted as $\beta$, is a scalar parameter that characterizes the interference of the modes in a waveguide. The WI establishes a relationship between frequency and range as acoustic waves propagate \cite{Chu:J82}. In particular, due to the modal interference, a broadband acoustic signal emitted from a moving source in shallow water will appear as striations, i.e., loci of almost constant intensities, in the $\rf$ surface plot. An $\rf$ surface plot is a spectrogram whose time-axis is converted to the source-receiver range. An example is shown in Fig.~\ref{fig:shipSoundEx}.

In a range-independent environment characterized by $\beta$, the striation referenced at a reference range and frequency ($r_0,f_0$) can be found at another frequency $f$ and the corresponding range $r$ that satisfies \cite[Ch. 2.4.6]{JenKupPorSch:B11}\vspace{-2mm}
\begin{equation}\label{eq:wi}
\frac{f}{f_0} = \biggl(\frac{r}{r_0}\biggr)^{\beta}\rmv\rmv. 
\end{equation}
This relationship has been utilized to estimate the range between mobile source and receiver \cite{CocSch:J10,ChoSonHod:J16} and considered for the navigation of autonomous underwater vehicles (AUVs) \cite{YouHarHicwRogKro:J20,JanMey:C23}. 


In this work, we consider a scenario in which a ship, moving on a straight path, passes by a submerged, stationary, single acoustic receiver in a range-independent, shallow-water environment. We propose a method for estimating the unknown range of the ship with respect to the receiver. The environment is assumed to be range- and azimuth-independent, i.e., it is characterized by a constant water depth and a sound speed profile that remains uniform in both range and azimuth.

The acoustic noise generated by a large moving vessel underwater arises from its machinery, propeller activity, and hydrodynamic interactions and can be described as a superposition of broadband components and a set of more intense narrowband tonal components \cite{McKRosWigHil:J12,ZhuGagMakRat:J22}. Young \etal \cite{YouHarHicwRogKro:J20} proposed a statistical method for the WI-based ranging when broadband components of the emitted sound are weak relative to the background noise. In that case, only the tonal components have significant signal-to-noise ratio (SNR), and the broadband component of the source signals can be ignored for the WI-based ranging. This paper considers an alternative, more common scenario where the emitted broadband components have significant SNR. In the considered dataset from the Seabed Characterization Experiment 2017 (SBCEX17) \cite{WilKnoNei:J20}, this assumption is valid for receiver ranges of more than $25$ km. Based on this dataset, we demonstrate that accurate ranging based on the WI is feasible using the broadband component of the received ship noise.

\section{Signal Model}
Consider an $\rf$ surface plot of the received ship noise that consists of $N$ snapshots. The time between snapshots is $t_{\Delta}$. Each snapshot $n \in \mathcal{N} =\{1,\dots,N\}$ 
corresponds to a range, $r_n$. There are $K$ frequency bins that contain only the broadband components of the noise emitted by the ship. The set of frequency bins with broadband components is denoted as $\Set{K}^{\text{b}}=\{1,...,K\}$. The received signal at range $r_n$ and frequency $f_k$, where $n \in \Set{N}$ and $k \in \Set{K}^{\text{b}}$, is modeled\vspace{-1mm} as
\begin{equation}\label{eq:general}
z_{n,k} = g_{n,k}s^{\symB}_{n,k} + u_{n,k},
\end{equation}
where $g_{n,k}\in\mathbb{C}$ is the deterministic channel transfer function, and $s^{\symB}_{n,k} \in \mathbb{C}$ and $u_{n,k} \in \mathbb{C}$ are random broadband source signal and background noise, respectively. Here, both the source signal and the noise are modeled by circularly symmetric complex Gaussian distributions, i.e., $s^{\symB}_{n,k} \sim \mathcal{CN}(0,(\varB_k)^2)$ and $u_{n,k} \sim \mathcal{CN}(0,(\varU_k)^2)$, with standard deviations $\varB_k$ and $\varU_k$, respectively. Note that the standard deviations are both frequency-dependent. 

We consider scenarios where, at the receiver, the contribution of the broadband source signal  is significantly higher than that of the background noise, i.e., $ (\varB_k)^2|g_{n,k}|^2 \gg (\varU_k)^2)$. Therefore, \eqref{eq:general} can be simplified\vspace{-.5mm} as
\begin{equation}
z_{n,k} = g_{n,k}s^{\symB}_{n,k}.
\nonumber
\end{equation}
As a result, the received acoustic signal $z_{n,k}$ follows a complex Gaussian PDF, i.e., $z_{n,k}\sim\mathcal{CN}(0,(\varB_k)^2|g_{n,k}|^2$). Its magnitude $x_{n,k} \rmv=\rmv |z_{n,k}|$ is Rayleigh distributed with the scale parameter $\theta_{n,k} = \varB_k|g_{n,k}|$, i.e., $x_{n,k} \sim \mathcal{R}\big(\theta_{n,k} \big)$. The magnitude $x_{n,k} $ is statistically independent across $n$ and $k$. Based on the functional form of $\mathcal{R}\big(\theta_{n,k} \big)$, the probability density function (PDF) of $x_{n,k}$ is given\vspace{-1mm} by
\begin{equation}\label{eq:rayleigh_ldf}
f_{\Set{R}}(x_{n,k};\theta_{n,k}) = \frac{x_{n,k}}{\theta_{n,k}^2}\exp{\Biggl(-\biggl(\frac{x_{n,k}}{2\theta_{n,k}}\biggr)^2\Biggr)}.
\end{equation}



The channel is modeled as follows. Without loss of generality, let $(n,k)$ and $(n',k')$, where $n,n'\in\Set{N}$ and $k,k'\in\Set{K}^{\text{b}}$, denote the range and frequency index pairs that, according to \eqref{eq:wi}, lie on the same striation. The primary assumption underlying the WI-based ranging is that the magnitude of the channel transfer function, commonly referred to as the Green's function, exhibits slow variation along the striations (see, e.g., \cite{SonByu:J20}). Based on this assumption, we model the channel transfer function \vspace{-1.5mm} as
\begin{equation}\label{eq:GScalarAssumption} 
|g_{n,k}| = \GScalar_{k} |g_{n',k'}|, 
\end{equation}
where the scaling factor $\GScalar_{k} \in \mathbb{R}^{+}$ is independent of the specific striation and depends only on the frequency. For future reference, $\M{Z}\in\mathbb{C}^{N\times K}$ is defined as the measurement matrix consisting of the measurement elements $z_{n,k}$, where $n\in\Set{N}$ and $k\in\Set{K}^{\text{b}}\vspace{-3mm}$.

\section{WI-based ranging}

For a given observed fixed measurement matrix $\M{Z}$, the joint likelihood function $\ell(r, \rr, \beta)$ depends on three unknown variables: the range $r$, the range rate vector $\rr = [\dot{r}_1, \dots, \dot{r}_{N-1}]^{\text{T}}\rmv\rmv$, and the WI $\beta$. The range $r = r_N$ corresponds to the last snapshot in the processed $\rf$ surface plot. Although the range rate vector $\rr$ does not explicitly appear in \eqref{eq:wi}, it is required to map the time in the spectrogram to range, a necessary preprocessing step for the application of \eqref{eq:wi}. Specifically, for a fixed reference range $r_N$, the range at a snapshot $n \in \mathcal{N}$ is expressed\vspace{-2.5mm} as
\begin{equation}\label{eq:rr}
r_i(r_n; \bm{\dot{r}}, t_\Delta) = r_n - \sum_{j=i}^{n-1} \dot{r}_j t_{\Delta}.
\end{equation}
Within the processing window, which comprises $N$ snapshots, the range rate is often assumed to remain constant. Consequently, $\rr$ can be replaced by a scalar range rate $\dot{r}$, where $\dot{r}_i = \dot{r}$ for $i \in \{1, \dots, N-1\}$. For simplicity, in what follows, a scalar range rate is assumed for the development of the proposed method. However, extending to the case where the range rate varies and a vector of range rate values is required to map time to range, as described in \eqref{eq:rr}, is straightforward.

The search intervals for $r$, $\dot{r}$, and $\beta$ are defined as $\Set{S}_{r} = [r_{\text{min}}, r_{\text{max}}]$, $\Set{S}_{\dot{r}} = [\dot{r}_{\text{min}}, \dot{r}_{\text{max}}]$, and $\Set{S}_{\beta} = [\beta_{\text{min}}, \beta_{\text{max}}]$, respectively. For convenience, we also introduce the joint search interval, i.e., $\Set{S} = \Set{S}_{r} \times \Set{S}_{\dot{r}} \times \Set{S}_{\beta}$, and the joint parameter vector is defined as $\V{q} = [r, \dot{r}, \beta]^{\text{T}}$. As discussed in detail in the journal paper of this work, due to ambiguities in the joint likelihood function, $\ell(\V{q})$, the estimation of any single parameter requires knowledge of the other two. For instance, performing ranging necessitates knowledge of $\dot{r}$ and $\beta$, meaning that the search intervals $\Set{S}_{\dot{r}}$ and $\Set{S}_{\beta}$ must be narrow.

\subsection{The WI and Range Rate}
The WI and range rate can be obtained using multiple approaches. In particular, in some scenarios, the range rate can be directly computed from the received acoustic field (see \cite{RakKup:J12, TaoHicKroKem:J07, SunGaoZhaGuoSonLi:J23}). In addition, if the frequency of tonal signal components around the closest point of approach (CPA) can be extracted, one can obtain the range rate from the Doppler shift\cite[Ch.8.4.2]{Bur:B02}. The WI parameter $\hat{\beta}$ characterizing the environment is either determined in a calibration step (see \cite{VerSarCorKup:J17}) or assumed to be 1 in shallow water \cite{JenKupPorSch:B11}.

\subsection{WI-Based Nonlinear Transformation}\label{sec:nlt}

\begin{figure*}[t]
   \raggedleft
   \begin{minipage}{0.325\textwidth}
      \centering
      \centerline{\includegraphics[width=\linewidth]{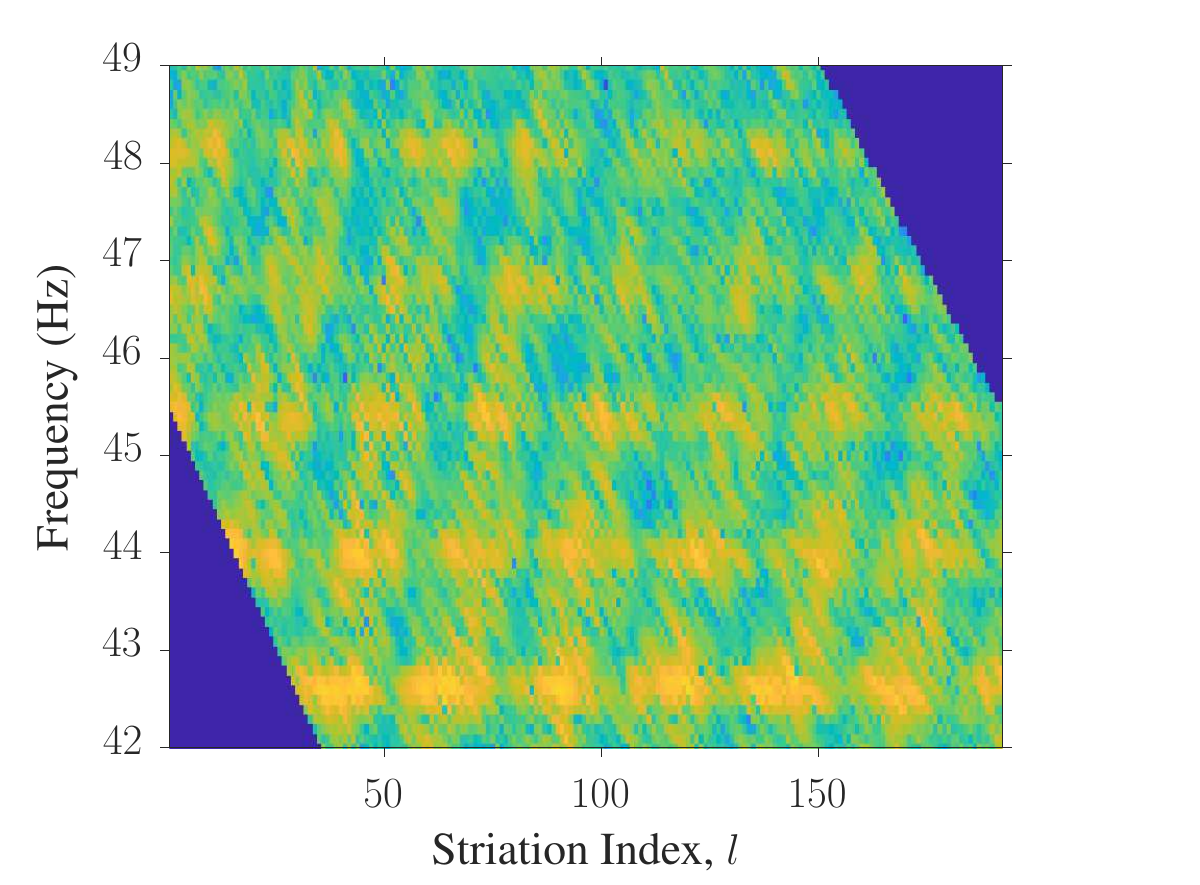}}
      \centerline{\scriptsize (a)}  
    \end{minipage}
   \begin{minipage}{0.325\textwidth}
      \centering
      \centerline{\includegraphics[width=\linewidth]{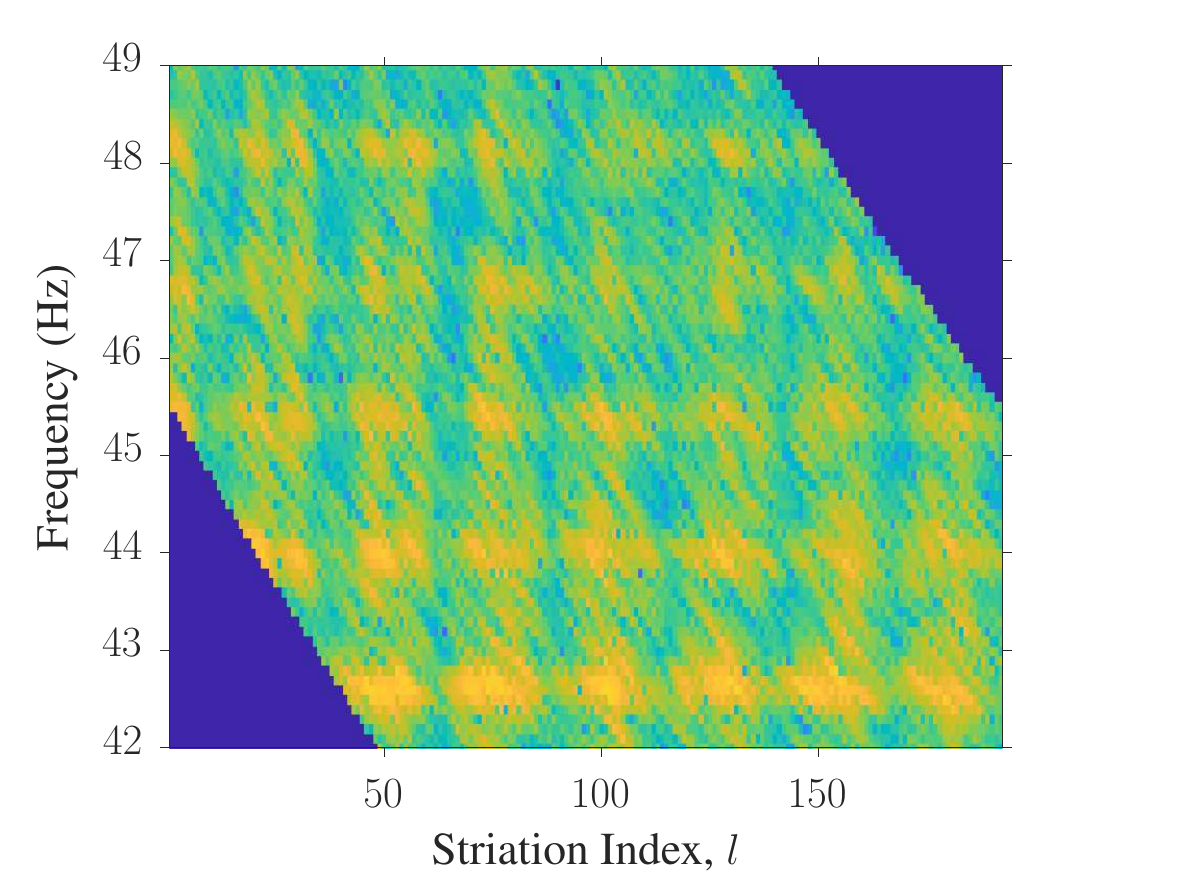}}
      \centerline{\scriptsize (b)}
   \end{minipage}
   \begin{minipage}{0.325\textwidth}
      \raggedright
      \centerline{\includegraphics[width=\linewidth]{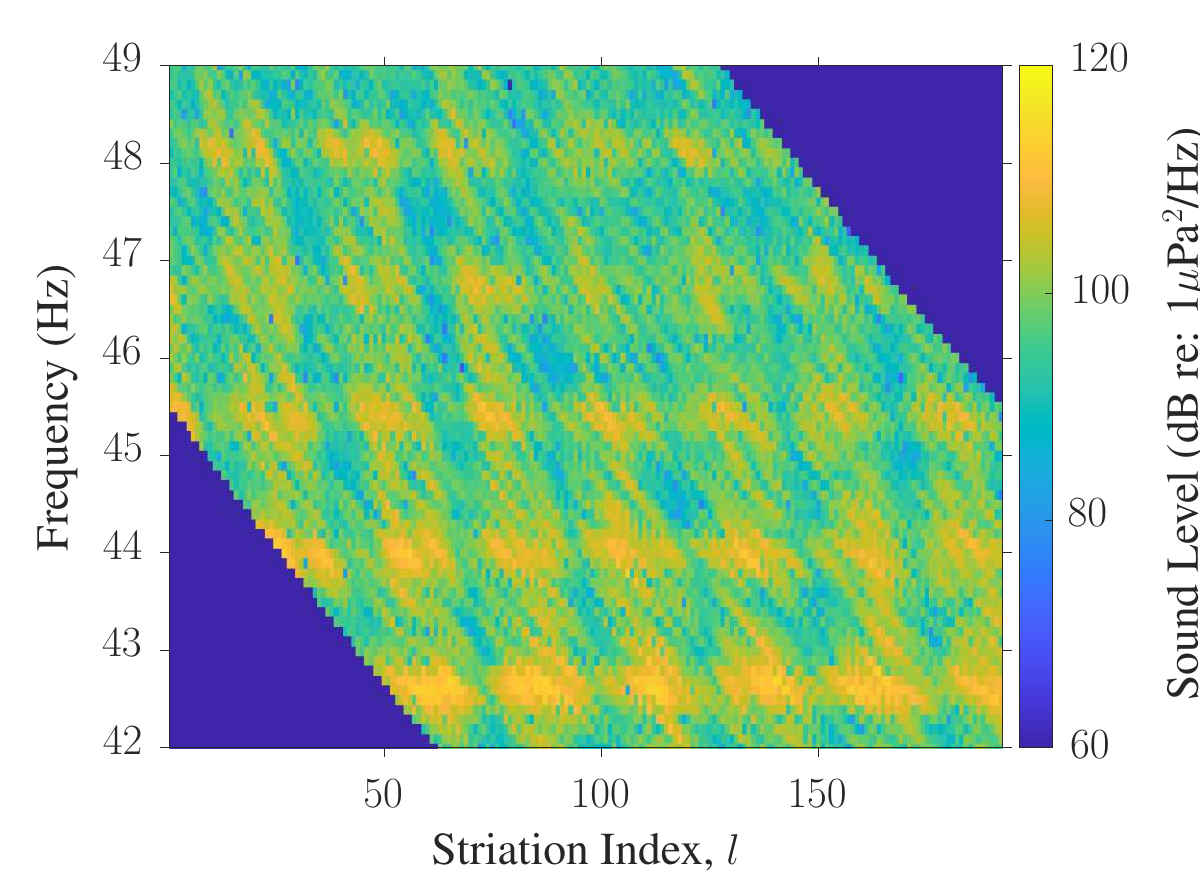}}
      \centerline{\scriptsize (c)}
   \end{minipage}
   \caption{
Examples of WI-based nonlinear transformations of the spectrogram from Fig.~\ref{fig:shipSoundEx}. The true range is $r = 22.7$ km. Hypothetical ranges are (a) $r = 17.7$ km, (b) $r_N = 22.7$ km, and (c) $r = 27.7$ km. Here, the range $r$ corresponds to the range at the last snapshot of the spectrogram, i.e., $r=r_N$. After transformation, each vertical sequence of snapshots forms a striation $l \in \SetStriation$. Transformation (b) achieves the highest joint likelihood, as intensities along each striation exhibit the most consistent levels. The parameters $\dot{r} = 10.2$ m/s (the constant range rate) and $\beta = 1.21$ (the WI) are used\vspace{-5mm}.} 
   \label{fig:NLT}
\end{figure*}

An essential step in evaluating the likelihood function $\ell(\V{q})$ is computing the acoustic fields and their magnitudes in the striation-frequency domain for a fixed parameter hypothesis $\V{q} = [r, \dot{r}, \beta]^{\text{T}}\in \Set{S}$. A striation is indexed by $l \in \Set{N}$. It adopts the index of the corresponding range bin $l \in \mathcal{N}$ at the reference frequency $f_{k'}$, where $k' \in \Set{K^\text{b}}$. At $f_{k'}$, the $l^\text{th}$ striation corresponds to the range-frequency pair $(r_{\spacel,{k'}}, f_{k'})$, where $r_{\spacel,k'} = r_l(r_N; \dot{r}, t_\Delta)$, as defined by \eqref{eq:rr}. For frequencies $f_k$, where $k\in \Set{K^{\text{b}}} \backslash \{ k'\}$, the range-frequency pair along the $l^{\text{th}}$ striation is $(r_{\spacel,k}, f_k)$, where $r_{\spacel,k}$ is obtained by using the nonlinear transformation obtained by rearranging ~\eqref{eq:wi}, i.e.\vspace{-1.5mm},
\begin{equation}\label{eq:wi2}
r_{\spacel,k} = r_{\spacel,k'} \Big(\frac{f_k}{f_{k'}}\Big)^{1/\beta}.
\vspace{-.5mm}
\end{equation}
We chose the reference frequency, $f_{k'}$, as the mid-frequency of the considered frequency band which led to the most accurate range estimates. This choice will be discussed in more detail in the journal version of this paper. 

Measurements in the striation-frequency domain are obtained by (i) mapping the time-axis of the spectrogram defined by measurement matrix $\M{Z}$ to a range-axis using \eqref{eq:rr}, and (ii) transforming the resulting the $\rf$ surface plot based on \eqref{eq:wi2} and the parameter hypothesis $\V{q}$. Examples of measurements in striation-frequency domain are shown in Fig.~\ref{fig:NLT}. Let us denote the measurement vector related to the $l^{\text{th}}$ striation as $\V{\underline{z}}_{\spacel}(\V{q}) = [\underline{z}_{\spacel,1}(\V{q}), \dots, \underline{z}_{\spacel,K}(\V{q})]^{\text{T}}$. After computing the new range values for $f_k$ using \eqref{eq:wi2}, the measurement $\underline{z}_{\spacel,k}(\V{q})$ is interpolated from the measurements $z_{n,k}$ on the original range grid $r_{n,k}$, $n \in \Set{N}$. Despite the interpolation, the statistics of $z_{n,k}$ are assumed for $\underline{z}_{\spacel,k}$ \cite{LaiKap:J22}. 

As shown in Fig.~\ref{fig:NLT}, after the nonlinear transformation, not all striations have valid intensity values for all frequencies $f_k$, where $k \in \Set{K}^{\text{b}}$. The number of striations with valid values for all frequencies depends on the used parameter hypothesis $\V{q}$. The minimum number of striations with valid values for all frequencies across all hypotheses $\V{q} \in \Set{S}$, is denoted as $M \leq N$. This value of $M$ is the number of striations with valid values for all frequencies related to hypothesis $\V{q}_{\Set{L}} = [r_{\text{max}}, \dot{r}_{\text{min}}, \beta_{\text{min}}]^{\text{T}}$. After $M$ has been determined by performing steps (i) and (ii), the set of striations used for the evaluation of $\ell(\V{q})$, can be indexed as $\SetStriation = \{N-M+1, \dots, N\} \subseteq \Set{N}$. In what follows, the underline notation (e.g., $\underline{z}$), indicates variables transformed using \eqref{eq:wi2}. This transformed variable depends on the parameter hypothesis used for the nonlinear transformation but, for simplicity, the dependency on $\V{q}$ is dropped where the context allows. After the transformation has been performed for a given $\V{q}$, the assumption in \eqref{eq:GScalarAssumption} can be written as $|\gTransformed_{\spacel,k}| = \GScalar_{k} |\gTransformed_{\spacel,k'}|$, $l \in \Set{L}$\vspace{-1mm}.

\section{Range Estimation}
\vspace{-.5mm}
In what follows, we present the method used for maximum likelihood (ML) estimation of the range, $r = r_N$, corresponding to the last, most recent, range in the $\rf$ surface plot, i.e., $\V{q} = [r, \dot{r}_{\text{true}}, \beta_{\text{true}}]^{\text{T}}\vspace{.5mm}$.

\subsection{Whitening of the Broadband Signal Components}\label{sec:broadWhiten}
The first step is to whiten the measurement matrix entries for all frequencies $f_k$ where $k \in \Set{K}^{\text{b}}$ with respect to a fixed, reference frequency $f_{k'}$ ($k'\in\Set{K}^{\text{b}}$). Let us define $\BScalar_{k} = \varB_k / \varB_{k'}$ as the ratio of the standard deviations at $f_k$ and $f_{k'}$ and let $\bm{s}^{\symB}_{n} = [s^{\symB}_{n,1},\dots,s^{\symB}_{n,K}]^{\text{T}}$ and $\bm{\BScalar} = [\BScalar_1,\dots,\BScalar_{K}]^{\text{T}}$ denote the broadband source signal vector and the standard deviation ratio vector, respectively. Note that $\BScalar_{k'} = 1$. The source signal vector $\bm{s}^{\symB}_{n} $ follows the PDF $\mathcal{CN}(\bm{0}, \text{diag}((\varB_{k'})^2 \bm{\BScalar}^2))$. 

Next, for a fixed reference range hypothesis $\V{q}$, consider the scale parameter $\underline{\theta}_{\spacel,k} = \varB_k |\gTransformed_{\spacel,k}|$ of $\underline{x}^{\symB}_{l,k} \sim \Set{R}(\underline{\theta}_{\spacel,k})$. Using the definitions of $\BScalar_{k}$ and \eqref{eq:GScalarAssumption}, we express $\underline{\theta}_{\spacel,k}$ as $\underline{\theta}_{\spacel,k} = \PScalarB_{k} \hspace{.3mm} \underline{\theta}_{\spacel,k'}$, where $\PScalarB_{k} = \eta_k \gamma_k$ is the ratio $\underline{\theta}_{\spacel,k} / \underline{\theta}_{\spacel,k'}$. To perform the whitening an estimate $\PScalarBScaHat_k$ of $\PScalarB_{k}$ is computed. This estimate relies on the fact that $\mathrm{Re}\{\underline{z}_{\hspace{.5mm} l,k'}\}, \mathrm{Im}\{\underline{z}_{\hspace{.5mm} l,k'}  \}\sim\mathcal{N}(0,\underline{\theta}_{\spacel,k'}^2)$ and $\mathrm{Re}\{\underline{z}_{\hspace{.5mm} l,k}\},\mathrm{Im}\{\underline{z}_{\hspace{.5mm} l,k}  \}\sim\mathcal{N}(0,\underline{\theta}_{\spacel,k}^2)$ with $\underline{\theta}_{\spacel,k} = \PScalarB_{k} \hspace{.3mm} \underline{\theta}_{\spacel,k'}$. Note that these four random variables are statistically independent. We define $2M$ independent and identically distributed random variables $v_{1,k} = \mathrm{Re}\{\underline{z}_{\hspace{.5mm} N-M+1,k}\}/\mathrm{Re}\{\underline{z}_{\hspace{.5mm} N-M+1,k'}\}$, $v_{2,k} =  \mathrm{Im}\{\underline{z}_{\hspace{.5mm} N-M+1,k}\}/\mathrm{Im}\{\underline{z}_{\hspace{.5mm} N-M+1,k'}\} $, \dots, $v_{2M-1,k} = \mathrm{Re}\{\underline{z}_{\hspace{.5mm} N,k}\}/\mathrm{Re}\{\underline{z}_{\hspace{.5mm} N,k'}\}$, $v_{2M,k} =  \mathrm{Im}\{\underline{z}_{\hspace{.5mm} N,k}\}/\mathrm{Im}\{\underline{z}_{\hspace{.5mm} N,k'}\}$. These random variables follow Cauchy distributions with location and scale parameters $0$ and $\PScalarB_{k}$, respectively. An estimate $\PScalarBScaHat_k$ of $\PScalarB_{k}$ is now obtained as half the sample interquartile range of these values \cite[Ch.7]{EvaHasPea:B00}. As a result, the whitened signal magnitudes for $l \in \Set{L}$ and $k \in \Set{K}^{\symB}$ are expressed as $\underline{x}^{\text{w}}_{\spacel,k} = \underline{x}_{\spacel,k} / \PScalarBScaHat_k$. Here, perfect whitening is assumed, implying whitened magnitudes $\underline{x}^{\text{w}}_{\spacel,k}\sim\Set{R}(\underline{\theta}_{\spacel})$, with $\underline{\theta}_{\spacel} = \underline{\theta}_{\spacel,k'}$.


\subsection{Rayleigh Scale Parameter Estimation Along a Striation}\label{sec:rayleighParam}
\begin{figure*}[t]
   \raggedleft
   \begin{minipage}{0.325\textwidth}
      \centering
      \centerline{\includegraphics[width=\linewidth]{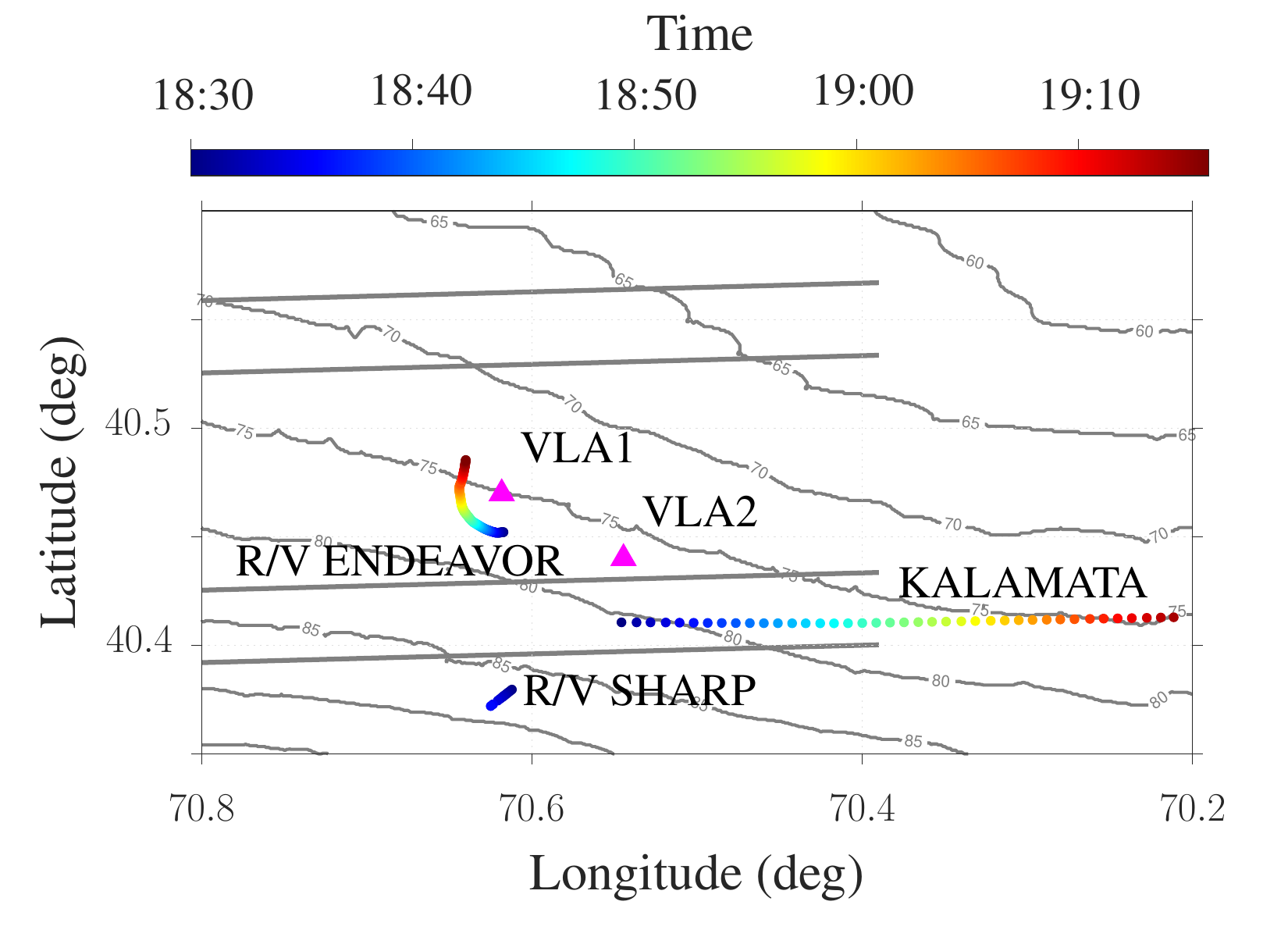}}
      \vspace{2mm}
      \centerline{\scriptsize (a)}  
    \end{minipage}
   \begin{minipage}{0.325\textwidth}
      \centering
      \centerline{\includegraphics[width=\linewidth]{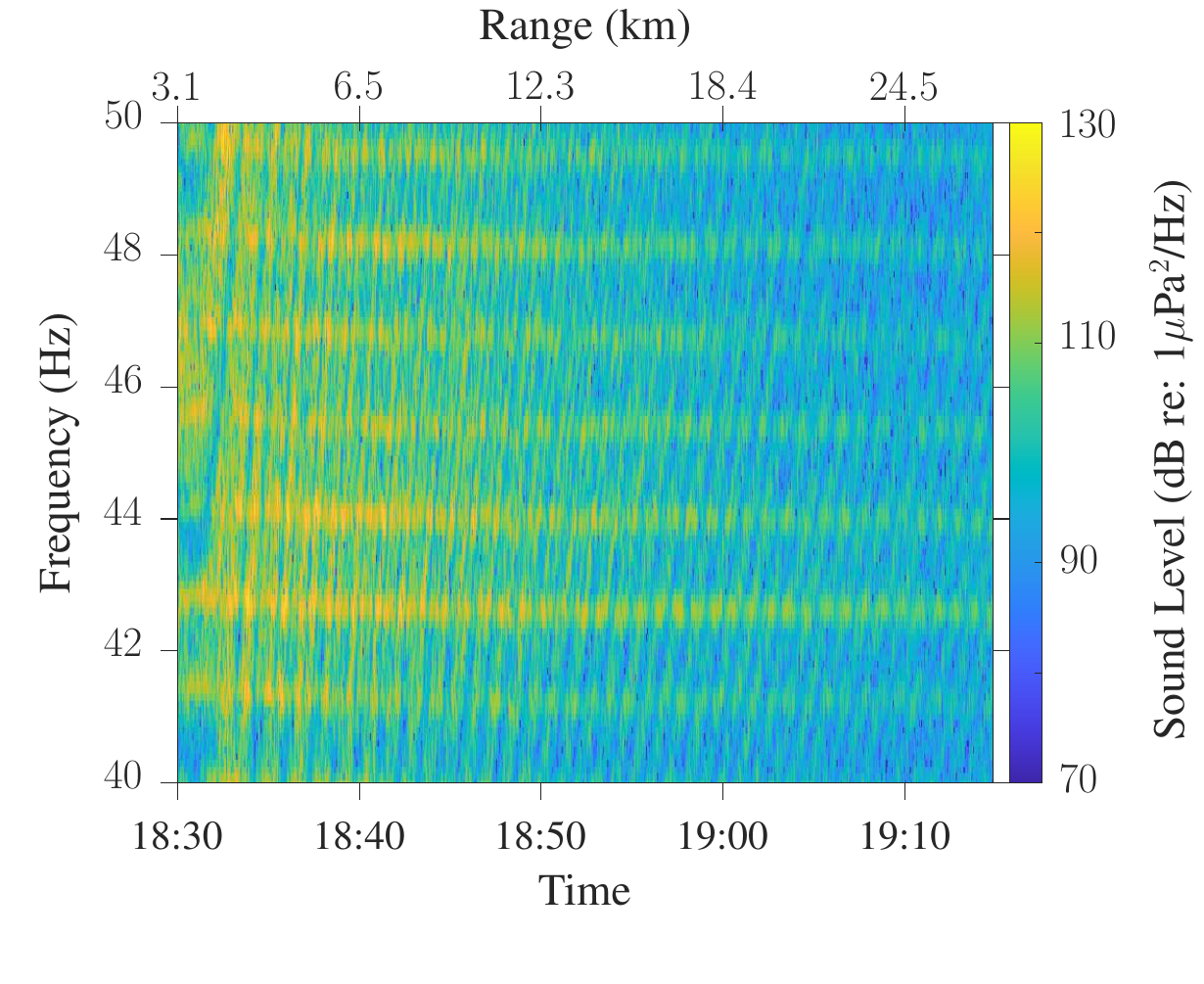}}
      \vspace{-3mm}
      \centerline{\scriptsize (b)}
   \end{minipage}
   \begin{minipage}{0.325\textwidth}
      \raggedright
      \centerline{\includegraphics[width=\linewidth]{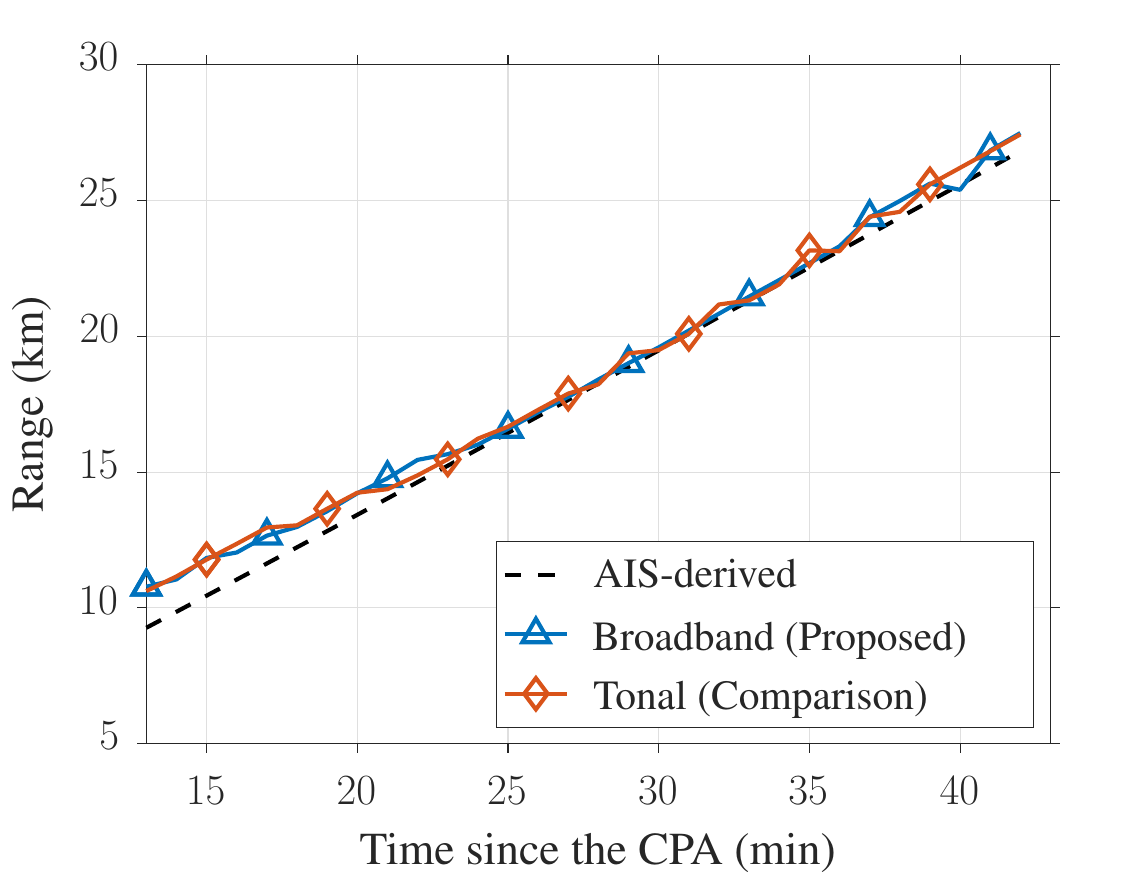}}
      \vspace{-1mm}
      \centerline{\scriptsize (c)}
   \end{minipage}
   \caption{(a) Considered scenario from SBCEX17. The track of the considered vessel KALAMATA between 18:30 and 19:15 on March 24th, 2017 (UTC) is shown. The tracks are located in the vicinity of the southern vertical line array (VLA2). The straight, gray lines indicate the shipping lanes. Two research vessels (Endeavor and Sharp) were operating in the vicinity, but their acoustic interference did not significantly affect our WI-based range estimation. (b) The spectrogram of the acoustic signal recorded during the considered event. Signals at frequencies $\Set{F}\in[42,49]$ Hz were processed. (c) The range estimation performance of the proposed method is compared with a reference method \cite{YouHarHicwRogKro:J20}. Ground truth range provided by AIS is also shown. Note that the comparison method relies on tonal signal components with high SNR, while the proposed method uses broadband signal components with lower SNR.\vspace{-5mm}} 
   \label{fig:r_result}
\end{figure*}

An ML estimate of $\underline{\theta}_{\spacel}$ can be obtained from the samples of $\underline{x}^{\text{w}}_{\hspace{.5mm} l,k}\hspace{0mm}$, $k \in \Set{K}^{\text{b}}$ as \cite[Ch.35]{EvaHasPea:B00}:
\begin{equation}
\hat{\underline{\theta}}_{\spacel} = \biggl(\frac{1}{2 K} \sum^{K}_{k=1}(\underline{x}^{\text{w}}_{\hspace{.5mm} l,k})^2\biggr)^{\hspace{-1mm} 1/2} \hspace{-1mm} . \nonumber
\end{equation}
In what follows, we use the notation $\underline{x}^{\text{w}}_{\hspace{.5mm} l,k}(\V{q})$ and $\hat{\underline{\theta}}_l(\V{q})$ to indicate that whitened samples and their scale parameter estimate are both functions of $\V{q}$. The vector\vspace{.3mm} of the estimated scale parameters is denoted as $\hat{\V{\underline{\theta}}} = [\hat{\underline{\theta}}_{N-M+1},\dots,\hat{\underline{\theta}}_{N}]^{\text{T}}$.

\subsection{Likelihood Function}\label{sec:BroadbandLikelihood}
Let $\underline{\M{X}}^{\text{w}}$ be the $M \times K$ matrix that consists of samples $\underline{x}^{\text{w}}_{\hspace{.5mm} l,k}(\V{q})\hspace{0mm}$ for $k \in \Set{K}^{\text{b}}$ obtain by a nonlinear transformation based on $\V{q}$. The proposed joint likelihood of $\V{q}$ is given by
\begin{align}\label{eq:jlb}
 \ell( \V{q} ) &= f^{\symB}(\underline{\M{X}}^{\text{w}} |\V{q})\nonumber\\[1.5mm]
&= \prod_{l\in\SetStriation}\prod_{k \in \Set{K}^{\symB}} f_{\Set{R}}\big(\underline{x}^{\text{w}}_{\spacel,k}(\V{q});\hat{\underline{\theta}}_{\spacel}(\V{q})\big). 
\end{align}
The functional form of the PDF $f_{\Set{R}}(\cdot)$ is provided in \eqref{eq:rayleigh_ldf}.

\subsection{ML Estimation}
Based on a joint likelihood function, $\ell( \V{q} )$, defined in \eqref{eq:jlb} and by setting $\V{q} = [r, \dot{r}_{\text{true}}, \beta_{\text{true}}]^{\text{T}}\vspace{.5mm},$ an ML estimate of range $r$, can be obtained\vspace{-.5mm} as 
\begin{equation}
\hat{r}^{\text{ML}} = \argmax_{r \in \mathcal{S}_r} \ell( r, \dot{r}_{\text{true}}, \beta_{\text{true}} ). \nonumber
\vspace{-1mm}
\end{equation}
Similarly, if the range and the range rate are known, one can use the same approach to obtain an ML estimate of the WI as $\hat{\beta}^{\text{ML}} = \argmax_{\beta \in \mathcal{S}_{\beta}} \ell( r_{\text{true}},\dot{r}_{\text{true}},\beta)$\vspace{-.5mm}.

\section{Data and Results}
\vspace{-.5mm}
To evaluate the proposed method, we analyzed a 42-minute acoustic recording from SBCEX17. The recording captured the acoustic signal of a container ship (KALAMATA, MMSI: 477510600) traveling in a straight line along a shipping lane at $10.2 \text{m}\ist\text{s}^{-1}$ southeast of vertical line hydrophone arrays (VLAs) deployed by the \textit{Scripps Institution of Oceanography}. The event began at 18:32 UTC on March 24, 2024, when the ship passed the closest point of approach (CPA = $3.1$ km) to the southern VLA ($40.442$\textdegree\ist N, $70.527$\textdegree\ist W). The seafloor depth ($75$ m) and sound speed profile (1473 $\text{m}\ist\text{s}^{-1}$) were nearly constant, making the environment relatively range-independent.

The proposed method's performance was compared to a modified version of the approach described in \cite{YouHarHicwRogKro:J20}, which relies on tonal components rather than broadband components. This approach assumes that broadband frequencies are primarily influenced by white background noise. The variances of this noise are estimated and used to calculate the instantaneous signal-to-noise ratio (SNR) for each tonal intensity. These instantaneous SNR values are modeled and employed as measurements for range estimation. Modifications to the comparison method were introduced to enhance computational efficiency and will be discussed in detail in a forthcoming journal publication.

The acoustic data were recorded by the eighth element of the VLA (33 m above the seafloor). Spectrograms were generated using 5-second segments with an additional 5 seconds of zero padding, 50\% overlap, and a Hamming window. The frequency band $\Set{F} \in [42, 49]$ Hz was analyzed, with tonal frequencies manually identified as $\Set{F}^{\ist\text{t}} = \{42.6, 44.0, 45.4, 46.7, 48.1\}$ Hz. The broadband frequencies were defined as $\FB = \{f\ist |\ist f \in \Set{F} ,\ist |f - f^{\text{t}}| \geq 0.4, f^{\text{t}} \in \Set{F}^{\text{t}} \}$ Hz. The proposed method processes $\Set{F}^{\text{b}}$, while the comparison method processes $\Set{F}^{\text{t}}$.

\subsubsection{WI Estimation}\label{sec:WIEstResult} The WI values were estimated every minute from 13 to 43 minutes after the CPA using the correct range rate calculated from the AIS. The estimation processed spectrograms of sufficient length $N$ to evaluate $183$ striations for all considered WI values in $\Set{S}_\beta=\{0.50,0.51,\dots,1.30\}$. Each spectrogram spanned approximately 12 to 15 minutes. The proposed and comparison methods yielded average WI estimates of $\hat{\beta}^{\text{b}} = 1.21$ and $\hat{\beta}^{\ist\text{t}} = 1.22$, both with a standard deviation of $0.03$, respectively. The comparison method assumes that the broadband frequencies contain pure noise, leading to a model mismatch since striations are observable in those frequencies. Additionally, the proposed method used 36 broadband frequency bins, compared to only five tonal bins for the comparison method, resulting in discrepancies in WI estimates.

\subsubsection{Range Estimation} Ranges between the ship and receiver were estimated every minute from 13 to 43 minutes after the CPA (18:45 - 19:15 UTC), using an assumed constant range rate of $10.2$ m\ist s\textsuperscript{-1}. The maximum observed range was $26.8$ km.  The proposed and comparison methods used their respective averaged WI estimates from Sec.~\ref{sec:WIEstResult}. Spectrogram lengths were chosen similarly, ensuring that $M=212$ striations were processed for all hypothetical ranges in $\Set{S}_r$. At each minute, the set of hypothetical ranges was defined to span $\pm$ 40\% of the range computed from the AIS, with 10-meter intervals. 

The range estimates were compared to the ground truth derived from AIS data (Fig.~\ref{fig:r_result} (c)). The root mean square errors of the range estimates were $662$ m for the proposed method and $686$ m for the comparison method, with corresponding error standard deviations of $433$ m and $446$ m, respectively. The large range estimates observed between minutes 13 and 20 after the CPA by both methods, relative to the ground truth, result from the assumed constant range rate being higher than the actual range rate near the CPA. This limitation can be addressed by using a vector range rate values in \eqref{eq:rr}, e.g., extracted from the Doppler shift of tonal components.

\section{Conclusion}
\vspace{-1mm}
In this paper, we introduced a method for ranging based on passive acoustic data that relies on waveguide invariant (WI) theory. Based on a new statistical model, our method can effectively estimate the value of the WI or the range of a large vessel with respect to a stationary, single acoustic receiver in shallow water. The proposed approach focuses on processing the broadband components of an acoustic signal emitted by the moving vessel due to its machinery, propeller activity, and hydrodynamic interactions. We demonstrated the capability to provide accurate range estimates based on real acoustic data. In particular, by processing data from SBCEX17, we validated the applicability of our approach in practical shallow-water environments. The presented results confirmed that the broadband component of the sound of large vessels emitted in a shallow water environment exhibits a spatial interference pattern that can be described by the WI and that this interference pattern can be effectively modeled to estimate both the WI and the range between the ship and receiver. 

Ongoing research explores statistical models that include both broadband and tonal components for the design of a more effective estimation method. In addition, the development of a method that can estimate range rate from the acoustic data used for the WI-based ranging would improve practicality and provide a more comprehensive ranging approach. This could, for instance, involve tracking the frequency of multiple tones \cite{GupKumBah:C13,MeyKroWilLauHlaBraWin:J18,JanMeySnyWigBauHil:J23} to extract Doppler shift information and range rate. Future work also includes comparisons with methods that integrate ray tracing and probabilistic data association \cite{WatStiTes:C24}.

\section*{Acknowledgment}
\vspace{-1mm}
We thank Dr. William Hodgkiss and David Ensberg for
providing the data from SBCEX17. 

This work was supported by the Defense Advanced Research Projects Agency (DARPA) under Grant D22AP00151 and by the Office of Naval Research (ONR) under Grant N00014-23-1-2284.

\renewcommand{\baselinestretch}{1.15}
\bibliographystyle{IEEEtran}
\bibliography{IEEEabrv,StringDefinitions,SALBooks,SALPapers,Temp}

\begin{thebibliography}{10}
\providecommand{\url}[1]{#1}
\csname url@samestyle\endcsname
\providecommand{\newblock}{\relax}
\providecommand{\bibinfo}[2]{#2}
\providecommand{\BIBentrySTDinterwordspacing}{\spaceskip=0pt\relax}
\providecommand{\BIBentryALTinterwordstretchfactor}{4}
\providecommand{\BIBentryALTinterwordspacing}{\spaceskip=\fontdimen2\font plus
\BIBentryALTinterwordstretchfactor\fontdimen3\font minus
  \fontdimen4\font\relax}
\providecommand{\BIBforeignlanguage}[2]{{%
\expandafter\ifx\csname l@#1\endcsname\relax
\typeout{** WARNING: IEEEtran.bst: No hyphenation pattern has been}%
\typeout{** loaded for the language `#1'. Using the pattern for}%
\typeout{** the default language instead.}%
\else
\language=\csname l@#1\endcsname
\fi
#2}}
\providecommand{\BIBdecl}{\relax}
\BIBdecl

\bibitem{Chu:J82}
S.~Chuprov, ``Interference structure of a sound field in a layered ocean,''
  \emph{Ocean Acoustics, Current State}, pp. 71--99, 1982.

\bibitem{JenKupPorSch:B11}
F.~B. Jensen, W.~A. Kuperman, M.~B. Porter, and H.~Schmidt, \emph{Computational
  Ocean Acoustics}, 2nd~ed.\hskip 1em plus 0.5em minus 0.4em\relax Springer
  Publishing Company, Incorporated, 2011.

\bibitem{CocSch:J10}
K.~L. Cockrell and H.~Schmidt, ``Robust passive range estimation using the
  waveguide invariant,'' \emph{J. Acoust. Soc. Am.}, vol. 127, no.~5, pp.
  2780--2789, 2010.

\bibitem{ChoSonHod:J16}
C.~Cho, H.~C. Song, and W.~S. Hodgkiss, ``Robust source-range estimation using
  the array/waveguide invariant and a vertical array,'' \emph{J. Acoust. Soc.
  Am.}, vol. 139, no.~1, pp. 63--69, 2016.

\bibitem{YouHarHicwRogKro:J20}
A.~H. Young, H.~A. Harms, G.~W. Hickman, J.~S. Rogers, and J.~L. Krolik,
  ``Waveguide-invariant-based ranging and receiver localization using tonal
  sources of opportunity,'' \emph{{IEEE} J. Ocean. Eng.}, vol.~45, no.~2, pp.
  631--644, 2020.

\bibitem{JanMey:C23}
J.~Jang and F.~Meyer, ``Navigation in shallow water using passive acoustic
  ranging,'' in \emph{Proc. Int. Conf. Inf. Fusion}, Charleston, SC, USA, 2023,
  pp. 1--8.

\bibitem{McKRosWigHil:J12}
M.~McKenna, D.~Ross, S.~Wiggins, and J.~Hildebrand, ``Underwater radiated noise
  from modern commercial ships,'' \emph{J. Acoust. Soc. Am.}, vol. 131, pp.
  92--103, 01 2012.

\bibitem{ZhuGagMakRat:J22}
C.~Zhu, T.~Gaggero, N.~C. Makris, and P.~Ratilal, ``Underwater sound
  characteristics of a ship with controllable pitch propeller,'' \emph{J. Mar.
  Sci. Eng.}, vol.~10, no.~3, 2022.

\bibitem{SonByu:J20}
H.~C. Song and G.~Byun, ``{Extrapolating Green's functions using the waveguide
  invariant theory},'' \emph{J. Acoust. Soc. Am.}, vol. 147, no.~4, pp.
  2150--2158, 04 2020.

\bibitem{RakKup:J12}
S.~T. Rakotonarivo and W.~A. Kuperman, ``Model-independent range localization
  of a moving source in shallow water,'' \emph{J. Acoust. Soc. Am.}, vol. 132,
  no.~4, pp. 2218--2223, 2012.

\bibitem{TaoHicKroKem:J07}
H.~Tao, G.~Hickman, J.~L. Krolik, and M.~Kemp, ``Single hydrophone passive
  localization of transiting acoustic sources,'' in \emph{Proc. IEEE OCEANS},
  2007, pp. 1--3.

\bibitem{SunGaoZhaGuoSonLi:J23}
K.~Sun, D.~Gao, X.~Zhao, D.~Guo, W.~Song, and Y.~Li, ``Estimation of target
  motion parameters from the tonal signals with a single hydrophone,''
  \emph{Sensors}, vol.~23, no.~15, 2023.

\bibitem{Bur:B02}
W.~Burdic, \emph{Underwater Acoustic System Analysis}, 2nd~ed.\hskip 1em plus
  0.5em minus 0.4em\relax Peninsula Publ., 2002.

\bibitem{VerSarCorKup:J17}
C.~M.~A. Verlinden, J.~Sarkar, B.~D. Cornuelle, and W.~A. Kuperman,
  ``{Determination of acoustic waveguide invariant using ships as sources of
  opportunity in a shallow water marine environment},'' \emph{J. Acoust. Soc.
  Am.}, vol. 141, no.~2, pp. 102--107, 02 2017.

\bibitem{LaiKap:J22}
L.~O. Lai and J.~O. Kaplan, ``A fast mean-preserving spline for interpolating
  interval data,'' \emph{J. Atmos. Ocean. Tech.}, vol.~39, no.~4, pp. 503 --
  512, 2022.

\bibitem{EvaHasPea:B00}
M.~Evans, N.~Hastings, and B.~Peacock, \emph{Statistical distributions},
  3rd~ed.\hskip 1em plus 0.5em minus 0.4em\relax John Wiley \& Sons, 2000.

\bibitem{WilKnoNei:J20}
P.~S. Wilson, D.~P. Knobles, and T.~B. Neilsen, ``Guest editorial an overview
  of the seabed characterization experiment,'' \emph{IEEE J. Ocean. Eng.},
  vol.~45, no.~1, pp. 1--13, 2020.

\bibitem{GupKumBah:C13}
R.~Gupta, A.~Kumar, and R.~Bahl, ``Motion parameter estimation of a radiating
  point source with multiple tonals using acoustic doppler analysis,'' in
  \emph{2013 IEEE OES Int. Symp. Underw}, 2013, pp. 1--6.

\bibitem{MeyKroWilLauHlaBraWin:J18}
F.~Meyer, T.~Kropfreiter, J.~L. Williams, R.~Lau, F.~Hlawatsch, P.~Braca, and
  M.~Z. Win, ``Message passing algorithms for scalable multitarget tracking,''
  \emph{Proc. {IEEE}}, vol. 106, no.~2, pp. 221--259, Feb. 2018.

\bibitem{JanMeySnyWigBauHil:J23}
J.~Jang, F.~Meyer, E.~R. Snyder, S.~M. Wiggins, S.~Baumann-Pickering, and J.~A.
  Hildebrand, ``{Bayesian detection and tracking of odontocetes in 3-D from
  their echolocation clicks},'' \emph{J. Acoust. Soc. Am.}, vol. 153, no.~5,
  pp. 2690--, 05 2023.

\bibitem{WatStiTes:C24}
L.~Watkins, P.~Stinco, A.~Tesei, and F.~Meyer, ``A probabilistic focalization
  approach for single receiver underwater localization,'' in \emph{Proc.
  FUSION-24}, 2024.

\end{thebibliography}

\end{document}